\documentclass[%
 reprint,  
 amsmath,amssymb,
 aip,
 jcp,
]{revtex4-2}
\usepackage[utf8]{inputenc}

\usepackage{cmap} 
\usepackage[T1]{fontenc}
\usepackage{lmodern} 
\usepackage{times}
\usepackage[scaled]{helvet}

\usepackage{amsmath,amsfonts}
\usepackage{mleftright} 
\usepackage{mathtools} 
\usepackage{siunitx} 

\usepackage{graphicx, grffile}
\newlength\figwidth
\usepackage{booktabs,dcolumn}
\newcolumntype{d}[1]{D{.}{.}{#1}}

\usepackage{url,hyperref} 
\usepackage[table,usenames,dvipsnames]{xcolor}
\hypersetup{colorlinks=true, linkcolor=BrickRed,
  urlcolor=blue!50!black, citecolor=blue!50!black}

\usepackage[capitalise]{cleveref}

\newcommand\bluesout{\bgroup\markoverwith{ \remark{\rule[0.5ex]{2pt}{0.4pt}}}\ULon}

\renewcommand\epsilon{\varepsilon}
\renewcommand\phi{\varphi}
\renewcommand\theta{\vartheta}
\renewcommand\rho{\varrho}
\renewcommand\vec[1]{{\boldsymbol #1}}
\newcommand\unitvec[1]{\vec{#1}}
\newcommand\tens[1]{\boldsymbol{#1}}
\newcommand\diff{\mathrm{d}}
\newcommand\Dt{\mathrm{D}_t}
\newcommand\e{\text{e}}

\renewcommand\geq\geqslant
\renewcommand\leq\leqslant

\newcommand\dslip{{\delta_\text{slip}}}
\newcommand\bslip{{\beta_\text{slip}}}
\newcommand\ddepl{{\delta_\text{depl}}}
\newcommand\eff{\text{eff}}
\newcommand\kB{k_{\text{B}}}

\begin{document}

\title{Fluid flow inside slit-shaped nanopores: the role of surface morphology at the molecular scale}

\newcommand\FUBaffiliation{\affiliation{Freie Universität Berlin, Institute of Mathematics, Arnimallee 6, 14195 Berlin, Germany}}
\newcommand\ZIBaffiliation{\affiliation{Zuse Institute Berlin, Takustraße 7, 14195 Berlin, Germany}}

\author{Giorgia Marcelli}
\FUBaffiliation

\author{Tecla Bottinelli Montandon}
\FUBaffiliation

\author{Roya Ebrahimi Viand}
\FUBaffiliation

\author{Felix Höfling}
\FUBaffiliation
\ZIBaffiliation
\email{f.hoefling@fu-berlin.de}

\date{\today}

\begin{abstract}
Non-equilibrium molecular dynamics (NEMD) simulations of fluid flow have highlighted the peculiarities of nanoscale flows
compared to classical fluid mechanics; in particular, boundary conditions can deviate from the no-slip behavior at macroscopic scales.
For fluid flow in slit-shaped nanopores, we demonstrate that surface morphology provides an efficient control on the slip length, which approaches zero when matching the molecular structures of the pore wall and the fluid.
Using boundary-driven, energy-conserving NEMD simulations with a pump-like driving mechanism, we examine two types of pore walls---mimicking a crystalline and an amorphous material---that exhibit markedly different surface resistances to flow.
The resulting flow velocity profiles are consistent with Poiseuille theory for incompressible, Newtonian fluids when adjusted for surface slip.
For the two pores, we observe partial slip and no-slip behavior, respectively.
The hydrodynamic permeability corroborates that the simulated flows are in the Darcy regime.
However, the confinement of the fluid gives rise to an effective viscosity below its bulk value;
wide pores exhibit a crossover between boundary and bulk-like flows.
Additionally, the thermal isolation of the flow causes a linear increase in fluid temperature along the flow,
which we relate to strong viscous dissipation and heat convection, utilizing conservation laws of fluid mechanics.
Noting that the investigated fluid model does not form droplets, our findings challenge the universality of previously reported correlations between slippage, solvophobicity, and a depletion zone.
Furthermore, they underscore the need for molecular-scale modeling to accurately capture the fluid dynamics near boundaries and in nanoporous materials, where macroscopic models may not be applicable.
\end{abstract}


\maketitle

\section{Introduction}

In the past 60 years, molecular dynamics (MD) simulations have developed into an indispensable tool in many areas of research, complementing experimental investigations and theoretical predictions; examples range from drug development \cite{mortier2015ddt} and cell biology \cite{enkavi2019cr} to materials research \cite{kuhn2014prb,mishin2015}, chemical process engineering \cite{bousige2016nm,ran_2024}, and geology \cite{zhong2024ggg}.
Nowadays, they are used routinely to study the physical and chemical properties of complex materials and macromolecules
as well as kinetic processes therein \cite{Vega_2008,kessler2015jpcb,piana2014cosb,schuette2023an}.
This provides detailed insights into phase diagrams, structural properties, transformation rates, and transport coefficients, including dynamic shear viscosity, thermal and electrical conductivity, and diffusivity.
The pioneering works \cite{alder1959jcp,rahman1964pr} employed classical particle models with simplistic interaction potentials to test fundamental predictions of statistical mechanics and kinetic theory.
In this respect, the results of MD simulations have set and are still setting corner stones in our understanding of material responses, which were supported by and are still challenging the theoretical developments, a prime example being the long-time tails in the autocorrelation of particles in liquids \cite{alder1967prl,ernst1971pla,straube2020cp,anikeenko2024jcp} and arrested fluids \cite{ernst1971pla2,hoefling2007prl,peng2016pre}.
Moreover, the development of advanced, realistic force fields \cite{Vega_2006,mishin2001prb,klauda2010jpcb,avendano2011jpcb}
has turned MD simulations into a prime method for quantitative predictions of material properties,
and these developments have rapidly gained new momentum with the advent of machine-learned interaction potentials that are based on first-principles calculations \cite{behler2007prl,bartok2015ijqc,zhang2018prl}.

Whereas the investigated system sizes were originally limited to a few hundred atoms and just several thousand integration time steps, simulations of some million atoms over tens of millions of time steps can be performed routinely on today's high-performance computing installations \cite{shibuta2015sr,mao2023ra,bouquiaux2023jcim,Schade2023ijhpc}, also owing to the massive parallelism provided by general-purpose accelerator hardware (``graphics processors'').
This has brought small-wave number and low-frequency phenomena within the reach of MD simulations   \cite{hoefling2015,roy2016,straube2020cp,chaudhuri2016prb,shibuta2015sr}, thereby giving access to hydrodynamic scales with molecular resolution.
It allows one to study set-ups typical of fluid mechanics in great detail such that also the boundary layers of a fluid flow can be resolved down to the scale of atomic interactions.
This is particularly relevant for the development of realistic continuum models of pores via up-scaling approaches, which need information about the molecular behavior at the nanoscale \cite{Blunt:COCIS2001,Tallarek:2022}; there, hydrodynamic permeability models are prone to break down \cite{Falk_2015}.

Transport phenomena in confined geometries, particularly at the nanoscale, showcase a continuously growing number of technological applications and relevance in natural processes \cite{Stein_2017, Li_2018, Phan_2020}.
The behavior of fluids in confined spaces is studied with the aim to design materials for filtration \cite{Koros2017,Wang2013}, energy applications \cite{Falk_2015, Wang2013, Malik_2019, Feng_2016}, water desalination \cite{Zhao_2015}, chip-level cooling \cite{lin2024pf}, and DNA sequencing \cite{Heerema_2016}, to name only a few.
A critical factor in determining fluid behavior within a porous medium is the nature of the interactions between the fluid and the pore surfaces \cite{sahimi_1993,Israelachvili2011}. Hydrophobicity, the tendency of a surface to repel water or similar polar fluids, modulates the fluid flow and distribution within the pore space;
in particular, it can lead to pronounced slippage of the flow at the fluid--wall interface.
Such slip, however, depends in a subtly way on factors such as surface roughness, wettability, and the presence of gaseous layers \cite{neto2005rpp}.
Surface roughness can change the fluid behavior significantly, which is known as the Lotus effect \cite{Israelachvili2011}: experiments evidence that specific surface coatings---such as polymers or nanotube structures, referred to as ``nanoforests''---can create superhydrophobic surfaces with extremely low friction to flow \cite{Lau2003}.
The effect has also been demonstrated in MD simulations of fluid flow near nano-structured surfaces \cite{cottin-bizonne2003nm}.
On the other hand, for hydrophilic surfaces with a similar topography, one may expect that trapping of fluids within surface asperities or pockets creates additional friction and reduces slippage.


For atomically smooth, crystalline surfaces, it was shown that hydro- and solvophobicity facilitate fluid slip \cite{bocquet_2010,huang2008prl}, which was linked to the formation of distinct interfacial layers, where fluid density and structure differ significantly from those in the bulk \cite{barrat1999prl}.
However, despite a general trend that hydrophobic surfaces show a reduced fluid--surface interaction and enhance fluid slip, hydrophobicity alone does not imply a large slip length and vice versa \cite{bocquet_2007}.
Moreover, subtle changes in the fluid--surface interaction can drive significant structural and density changes near the surface without modifying the wetting properties of the surface.
There is no simple correlation between the interfacial fluid structure and the hydrophobicity of the surface
and we refer to the partially controversial discussion in, \emph{e.g.}, Refs.~\citenum{evans2015jpcm,evans2017jcp,willard2014jcp}.
Also atomic-level details like surface composition, molecular orientation, and electronic interactions can profoundly impact the slip behavior, even on structurally similar surfaces, which was demonstrated in \emph{ab initio} simulations of water near graphene surfaces \cite{tocci2014nl,thiemann2022acs}.

Fluid flow through nanoporous media using non-equilibrium MD (NEMD) simulations has been investigated with low intensity since the 1990’s (\emph{e.g.}, Ref.~\citenum{Cracknell_1995}), but has seen a surge in research activity during the past decade. The focus is typically on the permeability of ordered porous media such as MOFs \cite{Velioglu:JMCA2019, Ozcan:CM2020} and linear or slit-shaped pores \cite{barrat1999prl, travis_2000, varnik2002jcp, cai_2008, frentrup_2012, FayazTorshizi:JPCB2022},
but also through disordered organic materials \cite{Botan:2013,Collell:2015,Falk_2015}. The majority of these studies addresses the transport of molecules in the gas phase or of low-density fluids, while NEMD simulations for dense liquids face particular technical challenges.
One approach to steady fluid flow is to prescribe the chemical potential of the inlet and outlet by means of grand-canonical (GC) reservoirs (dual control volume set-up) \cite{Heffelfinger:JCP1994,Botan:2013,Collell:2015}.
However, the Monte Carlo part of this approach would become inefficient for dense liquids due to the low acceptance rate of trial insertions of particles.
One solution to this problem is to control the pressure instead of the chemical potential \cite{ghoufi2011jcp},
and a recent, reservoir-based alternative to GC-NEMD simulations is offered by spatially adaptive resolution simulations (AdResS) \cite{Heidari:JCP2020,NoneqAdResS:JCP2020,EbrahimiViand:Thesis}.
Another line of approach to steady flow is driving the liquid in a boundary region of a periodic set-up by applying an external force to maintain a pressure difference \cite{frentrup_2012,FayazTorshizi:JPCB2022} or a concentration gradient \cite{Ozcan:CM2020}.
More generally, custom flow fields can be generated by extending the force field into the whole pore space \cite{renner2021prr};
in the context of incompressible flows, a constant force was used to mimic a constant pressure gradient \cite{barrat1999prl,varnik2002jcp}.
Below, we describe boundary-driven NEMD simulations of steady flow implemented via a pump-like driving mechanism, which is applied to the particles in a small region at the entrance of the pore, while preserving the Hamiltonian character of the fluid dynamics in the remaining pore volume.

No-slip boundary conditions are a customary and realistic choice in fluid engineering to model fluid flow through a macroscopic conduit.
On the other hand, the above discussion shows that the control of specific boundary conditions of fluid flow in MD simulations is a challenging task.
The ability to realize no-slip boundaries at the molecular scale would facilitate the comparison of particle-resolved simulations with predictions from continuum fluid dynamics.
In particular, one may expect that the latter predictions apply to a larger part of the simulated volume, which can serve as a coarse-grained particle model of fluid flow with low compressibility.
A possible route to this aim are surfaces with a molecular roughness in order to introduce tangential stresses on the fluid, and thereby increasing the friction with the surface and decreasing the slip.


In this work, we investigate how surface morphology, or molecular-scale surface topography, affects fluid flow near the surface.
In particular, how does it modify the effective boundary conditions, the local structure of the fluid, and the overall permeability?
We propose and characterize a nanopore model with discrete walls composed of obstacles arranged to match the molecular structure of the fluid.
We show that such a configuration allows efficient control of surface slip, which is almost completely eliminated.
More specifically, we simulate the steady flow of a simple fluid inside a slit-shaped nanopore as a minimal model of a porous medium; this choice enables comparison with previous studies and has obvious extensions to other pore cross sections.
We relate the molecular-level observations of fluid flow to the basic laws of fluid mechanics
and test predictions of Poiseuille theory for incompressible, Newtonian fluids.
Lastly, we discuss the issue of heat transport in isolated fluid flow under conditions of strong viscous dissipation, as encountered in fast-flow scenarios \cite{feng2012prl}.

\section{Simulation model}

\begin{figure}
    \centering
    \includegraphics[width=\figwidth]{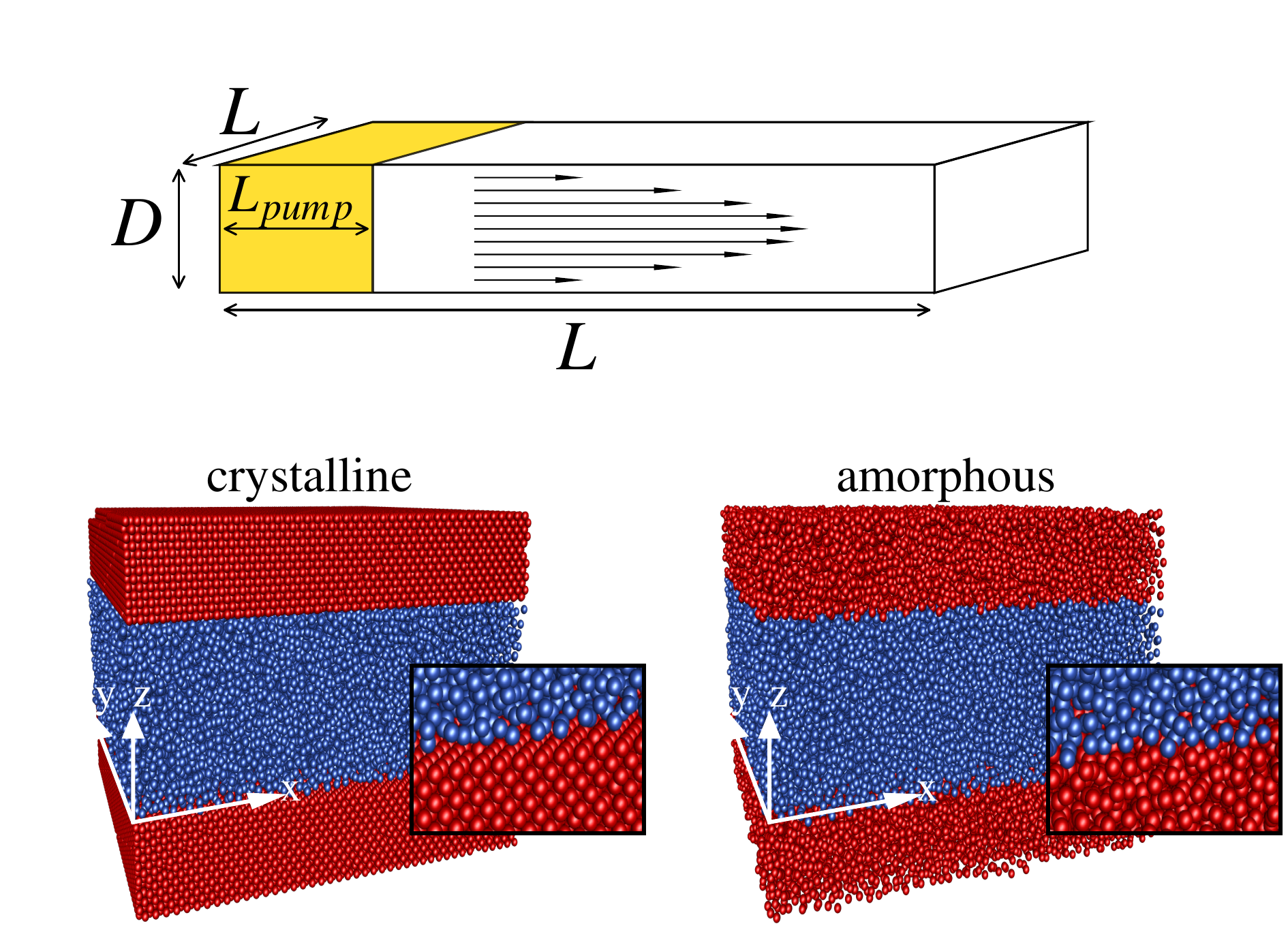}
    \caption{Top: sketch of the geometry of the simulated slit-shaped pore space. The cuboid is filled with fluid, which is driven at the pore entrance (yellow region) in the direction of the arrows, indicating the flow.
    The flow is delimited by the pore walls at the top and bottom faces of the cuboid, and periodic boundary conditions are applied at the other faces.
    ~Bottom: Snapshots of the particle configurations in NEMD simulations for two different obstacle arrangements (red particles), which realize walls with a crystalline (fcc) and amorphous structure, respectively.
    The pore is filled with fluid particles (blue).
    For better visibility, a smaller pore space ($L=50$) than the one used in the actual simulations is shown;
    the inset is a close-up of the fluid--wall interface.
    }
    \label{fig:wall_setup}
\end{figure}

We consider the fluid flow confined to a nanoscale conduit of slit geometry, which is created by two parallel, square-shaped walls of nominal distance $D$ and each of surface area $L\times L$, \cref{fig:wall_setup}.
The volume accessible to the fluid has the shape of a cuboid, of which two opposite open faces are referred to as the inlet and outlet of the conduit; periodic boundary conditions are used at the remaining two faces.
The application of a pressure difference $\Delta P=P_\text{in} - P_\text{out}$ between the inlet and outlet leads to plane Poiseuille flow; equivalently, one may impose a mean flow velocity on the fluid and observe the resulting pressure drop.

We choose a coordinate frame such that the $x$-axis points parallel to the wall surfaces, in the direction of the flow,
and the $z$-axis is normal to the surfaces, which are located at $z=\pm D/2$ so that $z=0$ denotes the plane in the center of the flow, \cref{fig:wall_setup}a).
By translational symmetry, the problem does not depend on the $y$-coordinate of the position.

\subsection{Fluid model}
\label{sec:fluid-model}

For the fluid, we employ a simple liquid consisting of repulsive spheres interacting via a smoothed Weeks--Chandler--Andersen (WCA) potential:
\begin{equation}
   U_\text{WCA}(r) =\left[U_\text{LJ}(r) - U_\text{LJ}(r_c)\right] \, g\boldsymbol((r-r_c)/h\boldsymbol),
   \label{eq:WCA}
\end{equation}
which is derived from the Lennard-Jones pair potential,
$
  U_\text{LJ}(r)=4\epsilon \bigl[(\sigma/r)^{12} - (\sigma/r)^6 \bigr].
$
The parameters $\epsilon$ and $\sigma$ serve as units of energy and length, respectively,
and the unit of time is chosen to be $\tau = \sqrt{m\sigma^2/\epsilon}$ in terms of the particle mass $m$.
Placing the cutoff distance $r_c$ at the minimum of the potential, $r_c=2^{1/6} \sigma$, the interaction contains only the repulsive forces that prevent particle overlap without considering attractive dispersion forces.
A smooth truncation is used to ensure good energy conservation during long MD simulations in the microcanonical (NVE) ensemble, implemented via multiplication of the energy-shifted potential with the algebraic interpolation function \cite{voigtmann2009,zausch2010,colberg2011,roy2016}
$g(x) = x^4 / \bigl(1+x^4\bigr)$ for $x \leq 0$ and $g(x) = 0$ otherwise.
Thereby, $U_\text{WCA}(r)$ is rendered $C^2$-continuous also at the cutoff.
We choose $h=0.005 \sigma$ for the truncation range, which is small enough to entail only minor changes of the fluid properties due to the truncation.
Due to the purely repulsive interaction, the fluid does not exhibit distinguishable liquid and vapor phases and, therefore, it does not form stable droplets in equilibrium.

Throughout this study, density and temperature of the quiescent fluid, without driving, are fixed to
$\rho = 0.7\sigma^{-3}$ and $T= 1 \epsilon/\kB$.
We focus on a pore width of $D=15\sigma$, amounting to $N=\num{105000}$ fluid particles in the set up, which is complemented by results for wider pores with $D=30\sigma$ and $60\sigma$ to better separate surface and confinement effects.
For later reference, some material constants and transport coefficients of the WCA fluid are given in \cref{tab:bulk-properties} at a selected thermodynamic state points $(\rho, T)$ representative of the flowing fluid.

\begin{table*}
\footnotesize\tabcolsep=1em
\begin{tabular}{rr@{$\:=\:$}rr@{$\quad$}rr@{$\quad$}ll}
\toprule
\toprule
temperature & $T$ & 1.0 & 1.0 & 3.0 & $\epsilon / \kB $ \\
mass density & $m\rho$ & 0.7 & 0.9 & 0.55 & $m \sigma^{-3}$ \\
pressure & $P$ & 4.277 & 10.130 & 5.360 & $\epsilon \sigma^{-3}$\\
internal energy & $e$ & $1.462(2)$ & $1.334(5)$  & $4.419(3)$ &$\epsilon / m$ \\
\midrule
isochoric specific heat & $c_V$ & $2.034(1)$ & $2.527(5)$ & $1.791(1)$ & $\kB / m$ \\
compressibility & $\chi_T$ & $0.04256(1
)$ & $0.00991(2) $ & $0.0600(1) $ & $\sigma^3 / \epsilon$ \\
speed of sound & $c_s$ & $5.79(1)$ & $10.58(1)$ &  $5.50(1)$ &$\sigma/\tau$ \\
shear viscosity & $\eta_b$ & $1.423(2)$ & $3.494(6)$ & $4.17(2)$ & $\sigma^3 \tau / \epsilon$ \\
thermal conductivity & $\kappa$ & $9.38(3)$ & $13.51(5)$ & $9.15(4)$ & $\kB / (\sigma \tau)$ \\
\bottomrule
\bottomrule
\end{tabular}
\caption{
Thermodynamic observables, material constants, and transport coefficients of the purely repulsive, WCA fluid, at three thermodynamic state points $(\rho, T)$ representative of the flowing fluid;
the quiescent fluid corresponds to $T=1.0 \epsilon/\kB$ and $\rho=0.7 \sigma^{-3}$.
The values were obtained with equilibrium MD simulations following standard procedures, the speed of sounds
was calculated from $c_s = (\rho \chi_T)^{-1/2}$.
Numbers in parentheses give the uncertainty in the last digit.
}
\label{tab:bulk-properties}
\end{table*}

\subsection{Pore morphologies}
\label{sec:wall-setup}


The pore walls are implemented as arrangements of obstacle particles, which are fixed in space and which exert forces on the fluid particles according to the same WCA potential, \cref{eq:WCA}, that is used for the interaction between fluid particles.
The use of such discrete walls provides us with flexibility regarding the surface morphology.
This allows us to study the role of morphology in the fluid flow by comparing simulations with two types of pore surfaces, which we refer to as \emph{crystalline} and as \emph{amorphous} walls, \cref{fig:wall_setup}.


For crystalline walls, the pore particles are placed on a face-centered cubic (fcc) lattice with a lattice constant of $a_\text{lat}=1.15 \sigma$ such that the lattice planes are located at $z=\pm[D/2 + (n - 1/2) a_\text{lat}/2]$ for integer $n \geq 1$.
Technically, the dimension of the simulation box is extended along the $z$-axis by adding $15\sigma$ in total to accommodate the obstacle particles.
The lattice constant is chosen small enough to prevent infiltration of fluid particles into the wall structure,
it implies nearest-neighbor distances of $a_\text{lat}/\sqrt{2} \approx 0.82 \sigma$.
The dense arrangement of pore particles results in a highly ordered, crystalline structure, exposing a comparably smooth surface to the fluid.


For amorphous walls, the pore particles are distributed in a disordered fashion in such a way that the fluid structure is matched;
a similar approach was used earlier to study confinement effects in supercooled liquids \cite{scheidler2000epl,Scheidler2004}.
Such wall structures are generated by filling the whole simulation volume, $(D+15\sigma)\times L \times L$, with a WCA fluid at the desired fluid density and temperature and equilibrating its structure in a short NVT simulation run.
Afterwards, the particles found outside the (nominal) volume of the conduit are identified as obstacles and they delimit the pore space, while those inside the pore region represent the fluid; obstacle particles do not participate in the further integration of the equations of motion.
The procedure leads to comparably rough, irregular wall surfaces in contact with the fluid.

Due to the presence of density fluctuations in the obstacle configuration and the overall lower obstacle density,
a small number of fluid particles can penetrate deep into the pore walls and get ``lost''.
As an additional measure to ensure particle conservation in the pore volume, we placed two flat, impenetrable walls alongside the obstacles at positions $z_\text{flat}^{\pm}=\pm(D/2 + \sigma)$, inside of the pore walls.
For the interaction energy of a fluid particle at position $z$ with each wall, we use a purely repulsive surface potential:
\begin{equation}
    U_\text{flat}(\hat z < r_c) = \frac{4\pi\epsilon_w}{3\times 15}
        \Bigl(\frac{\sigma}{\hat z}\Bigr)^9,
    \label{eq:planar_wall}
\end{equation}
in terms of the distances $\hat z =z-z_\text{flat}^{+}$ and $z=z-z_\text{flat}^{-}$.
The surface potential is smoothly truncated, analogously to \cref{eq:WCA}, using again the cutoff distance $r_c=2^{1/6} \sigma$ with $h=0.005\sigma$.
We emphasize that these additional, flat potentials are positioned at a distance of $\sigma$ well behind the pore surfaces and that the particles of the moving fluid are essentially not interacting with them.

\section{Fluid mechanical considerations}

\subsection{Poiseuille flow in a slit-shaped conduit}
\label{Sec:HP}

In fluid mechanics, the laminar flow through a slit-like conduit is referred to as plane Poiseuille flow \cite{Kundu:FluidMechanics, Landau2013}, which develops in response to an applied pressure difference $\Delta P$ along the conduit. The latter is of length $L$ and width $D$ and the extension in the remaining direction (along the $y$-axis) is taken to be macroscopically large (alternatively, periodic boundary conditions can be applied, as we have done in the NEMD simulations). Here, we shall briefly recall the main results of the theory with an emphasis on the underlying assumptions.

The steady flow of a uniform, incompressible, Newtonian fluid at low Reynolds number is governed by the stationary Stokes equation for the velocity field $\vec v(x,z)$ and the pressure field $P(x,z)$.
In the absence of an external force and far from the pore entrance, the problem is $x$-independent and the flow is fully developed and aligned with the axis of the conduit, $\vec v = (v_x, 0, 0)$.
For low Mach number, due to the incompressibility, the pressure gradient is aligned with the flow and thus constant, $\partial_x P(x,z)= -\Delta P/L$. This simplifies the stationary Stokes equation to read:
\begin{equation}
0 = \Delta P / L + \eta \partial^2_z v_x(z),
\label{eq:PDE-HP}
\end{equation}
where $\eta$ is the dynamic shear viscosity. In Poiseuille theory, $\eta$ is assumed to be a constant parameter.

At macroscopic scales, no-slip boundary conditions hold at the surfaces of the conduit,
$\vec v(x,\pm D/2) = 0$.
At the atomistic scale, the friction near surfaces is not fully developed and fluid molecules can slip along the surface. The arrangement of surface atoms and also their interaction with the fluid molecules affect the boundary layer of the flow. These lead to an effective full or partial slip of the flow field, described by Navier's partial slip condition \cite{navier1823mouvement, Landau2013}:
\begin{equation}
    v_x(z = \pm D/2) = \mp \bslip \,\partial_z v_x(z=\pm D/2),
    \label{eq:partial-slip}
\end{equation}
where $\bslip$ is the slip length, representing the distance below the surface at which the no-slip condition would hold if the velocity was linearly extrapolated into the wall.
As an alternative, one may use a parabolic extrapolation of the flow profile near the surface, which yields a length $\dslip$ that is interpreted as an apparent displacement of the pore surface from, e.g., $z=D/2$ to $z=D/2+\dslip$.
Both lengths are related via:
\begin{equation}
    \bslip = \dslip (1 + \dslip / D) \,.
\end{equation}
Solving \cref{eq:PDE-HP} for the boundary condition \eqref{eq:partial-slip} leads to the velocity profile of Poiseuille flow:
\begin{equation}
    v_x(z) = \frac{\Delta P}{2 \eta L} \left[\left(\frac{D}{2} + \dslip\right)^2 - z^2 \right].
    \label{eq:HP-slip}
\end{equation}
The no-slip solution is contained for $\dslip = \bslip = 0$.

\subsection{Permeability}

In many applications, interest often lies in the total amount of fluid passing through the conduit per unit time, which is quantified by the \emph{volumetric flow rate} or, equivalently, the \emph{mass flux} per unit area,
\begin{equation}
  J_\rho = \frac{1}{D} \int_{-D/2}^{D/2} m \rho v_x(z) \, \diff z \,,
  \label{eq:mass-flux-def-a}
\end{equation}
given as the momentum density field, $m \rho \vec v$, averaged over the cross section of the conduit.
The fluid flow is proportional to the applied pressure gradient, if the latter is not too large. This is known as Darcy's law in the field of porous media \cite{MUSKAT_1938, sahimi_1993}, which defines the permeability $k$ of the medium:
\begin{equation}
  J_\rho = \frac{k \Delta P}{\nu L}
  \label{eq:Darcy-a}
\end{equation}
in terms of the kinematic viscosity $\nu = \eta / (m \rho)$.
For Poiseuille flow, \cref{eq:HP-slip}, one finds for the permeability of the slit-shaped conduit:
\begin{align}
     k &=\frac{D^2}{12}+\bslip \,\frac{D}{2} \,.
    \label{eq:permeability_slip}
\end{align}
Thus, the presence of slip increases the effective permeability of the pore, allowing more fluid to pass through for the same pressure gradient, which is intuitively expected.

\subsection{Non-isothermal flow}
\label{sub:non-isothermal-flow}

For the interpretation of data obtained within energy-conserving MD simulations of energetically isolated flows, as presented below, we need to relate viscous dissipation and heat conduction to the fluid temperature.
Such a relation can be found in selected fluid mechanics texts \cite{Kundu:FluidMechanics,winter1987};
we summarize it here and sketch its derivation in \cref{sec:temperature-equation}.

Combining the fundamental conservation laws of mass, momentum, and energy with the caloric equation of state of the fluid, under the assumption of local thermodynamic equilibrium and the absence of external forces, one obtains the energy equation in its form for the temperature field~$T$:
\begin{equation}
    c_v m\rho (\partial_t +\vec v \cdot \nabla) T = \dot q + \Phi \,,
    \label{eq:temperature-eq}
\end{equation}
where $c_v$ is the isochoric specific heat, $q$ denotes the irreversibly added heat density field (per volume),
and the viscous dissipation function $\Phi$ describes the heat that is locally introduced into the fluid due to viscous stresses as the fluid moves non-uniformly.
\Cref{eq:temperature-eq} rests on mild assumptions and, in particular, holds for non-uniform, compressible, non-Newtonian fluids. It remains to specify the terms of the \emph{r.h.s.} of the equation.

Assuming that the heat density of the quiescent fluid changes only due to heat conduction and that the temperature gradient is sufficiently small, the heat flux obeys Fourier's law and the rate of change of the heat density follows to be
$\dot q  = \kappa \nabla^2 T$, \cref{eq:heat-equation-a};
the coefficient $\kappa$ is the thermal conductivity.
Specializing further to incompressible, Newtonian fluids, the viscous dissipation function can be expressed in terms of the
rate of strain tensor $\dot{\tens\gamma} = \nabla \vec v + (\nabla \vec v)^\top$
and reads: $\Phi = \eta \dot{\tens\gamma} : \dot{\tens\gamma} / 2$, \cref{eq:Phi-incompressible-a}.
Substituting into \cref{eq:temperature-eq} yields to the equation of the temperature field under the named approximations \cite{winter1987}:
\begin{equation}
    c_v m\rho (\partial_t T + \vec v \cdot \nabla T) = \kappa \nabla^2 T + \tfrac{1}{2} \eta \dot{\tens\gamma} : \dot{\tens\gamma} \,.
    \label{eq:temperature-eq-viscous}
\end{equation}
It describes how the local temperature of the fluid changes as a result of fluid convection ($\vec v \cdot \nabla T$), heat conduction within the fluid ($\kappa \nabla^2 T$), and viscous dissipation ($\eta \dot{\tens\gamma} : \dot{\tens\gamma} / 2$); the latter accounts for the irreversible work performed due to friction between fluid layers.

\Cref{eq:temperature-eq-viscous} complements and couples to the Stokes equation, \cref{eq:PDE-HP}.
Focusing on steady flow in the above slit geometry, \cref{eq:temperature-eq-viscous} reads:
\begin{equation}
    c_v m \rho v_x \partial_x T
    = \kappa (\partial_x^2 T +\partial_z^2 T) + \Phi,
    \label{eq:energyeqVII}
\end{equation}
where the viscous dissipation function of uni-directional, incompressible Newtonian flow is given by
$\Phi(z) = \eta (\partial_z v_x)^2$.
Using \cref{eq:PDE-HP} and the symmetry at $z=0$, it follows that
\begin{equation}
   \Phi(z) = \frac{(\partial_x P)^2}{\eta} \, z^2 \,,
   \label{eq:viscous-dissipation}
\end{equation}
which is independent of the boundary conditions at the surfaces, also see \cref{eq:HP-slip}.
For simplicity, we will treat the shear viscosity $\eta$ as being uniform in space, which needs to be justified on a case-by-case basis. Specifically, we neglect the dependencies of $\eta$ on temperature, density, and shear rate,
noting that a more advanced, even non-local theory may be more realistic \cite{cadusch2008jpa}.
To make further progress, we assume that the temperature is transversally equilibrated, \emph{i.e.}, in the pore cross section, so that we can approximate $\partial_z T(x,z) \approx \text{const}$ and set $\bar T(x) := T(x,z)$.
With this, the mass flux $J_\rho$ appears on the \emph{l.h.s.}\ of \cref{eq:energyeqVII} after integration over $z$;
due to mass conservation, \cref{eq:mass-cons-a}, $J_\rho$ through any cross section of the pore is constant along the steady flow, \emph{i.e.}, it does not depend on $x$.
In addition, we introduce the transversally averaged viscous dissipation:
\begin{align}
    \bar \Phi &= \frac{1}{D} \int_{-D/2}^{D/2} \Phi(z) \diff z = \frac{(\partial_x P)^2}{\eta} \frac{D^2}{12}.
    \label{eq:Phi-bar}
\end{align}
Together, this leads to an inhomogeneous, first-order ordinary differential equation in $\partial_x \bar T(x)$:
\begin{equation}
    \kappa \partial_x^2 \bar T(x) - c_v J_\rho \partial_x \bar T(x) = -\bar \Phi.
    \label{eq:temperature-ode}
\end{equation}
A particular solution is given by the constant temperature gradient:
\begin{equation}
    \partial_x \bar T = \frac{\bar \Phi}{c_v J_\rho} \,,
    \label{eq:grad-T}
\end{equation}
which implies a linear increase of the temperature $\bar T(x)$ along the flow.
For Poiseuille flow, this simplifies to:
\begin{equation}
    \partial_x \bar T = \frac{\partial_x P}{c_v m \rho} \,\frac{1}{1+6\bslip / D} \,,
    \label{eq:grad-T-Poiseuille}
\end{equation}
where we have expressed $J_\rho$ via Darcy's law and the permeability of the slit pore, \cref{eq:Darcy-a,eq:permeability_slip}.

The general solution to \cref{eq:temperature-ode} exhibits an exponential variation of $\partial_x \bar T$ with characteristic length
$\ell := \kappa / (c_v J_\rho)$:
\begin{equation}
    \partial_x \bar T(x) = \frac{\bar \Phi}{c_v J_\rho} + \left(\bar T'_0 - \frac{\bar \Phi}{c_v J_\rho} \right) \,\e^{x/\ell}
    \label{eq:grad-T-general}
\end{equation}
for arbitrary initial value $\bar T'_0 := \partial_x \bar T(0)$.
We note that this unstable behavior does not play a role in the following interpretation of the present NEMD simulations.
However, it is likely to occur in the absence of a pressure gradient, where $\bar\Phi = 0$ due to \cref{eq:Phi-bar};
an exponential increase of the fluid temperature was observed in simulations of the frictionless bulk flow in front of a constriction or porous medium \cite{EbrahimiViand:Thesis}.

\section{Simulation method}

\subsection{Boundary-driven NEMD simulation}

\paragraph*{Thermostatted pump.}

Boundary-driven flows are generated within the NEMD simulations by placing a ``pump'', combined with a thermostat, at the entrance of the pore space \cite{EbrahimiViand:Thesis}.
For particles in the region $0 \leq x \leq L_\text{pump}=10 \sigma$ along the $x$-axis, new velocities are randomly assigned, after fixed intervals of $\tau_\text{pump} = 0.2\tau$, by sampling from a shifted Gaussian distribution with mean $v_0 \vec e_x$ and variance $\kB T/m$; the latter corresponds to the Maxwell--Boltzmann distribution at temperature $T$ with an imposed flow velocity $v_0$ along the $x$-axis.
Thereby, particles leave the pump region with the desired target velocity, pushing away particles in front of them and thereby creating a steady mass flux in the pore.
We note that due to the resistance of the pore to the flow, the achieved average flow velocity is slightly below the nominal pump velocity $v_0$ (see below).
This issue can be mitigated by decreasing the update interval $\tau_\text{pump}$.
For the data interpretation, we will work with the achieved flow velocity.

The motion of all fluid particles, including those in the pump region, is generated with the velocity Verlet integrator;
the fluid inside the pump is additionally forced to a prescribed temperature and velocity.
This yields a Hamiltonian dynamics in the remaining pore space ($L_\text{pump} < x < L$), where mass, momentum, and energy are thus locally conserved.
The integration timestep was chosen to be $\delta t=0.001\tau$ for both equilibrium MD and NEMD simulations.

\paragraph*{Simulation protocol.}

The system preparation starts with an equilibration phase in a canonical (NVT) ensemble over a duration of $10^3 \tau$, where the number of particles and the volume of the simulation box are fixed and a global Andersen thermostat \cite{frenkel_2001}, with an update rate of $10\tau^{-1}$, is used to control the temperature.
After equilibration, the simulation transitions to the non-equilibrium set-up, by replacing the Andersen thermostat with the thermostatted pump in the boundary region.
Two NEMD simulation runs follow, each of a duration of $10^3\tau$, first to relax the fluid to the non-equilibrium steady state and, finally, to collect the data for the observables.
The used relaxation time is much longer than the time needed to establish the pressure gradient along the conduit, which we estimate from the speed of sound $c_s$ to be of the order of $L/c_s \lesssim 20\tau$, see \cref{tab:bulk-properties}.
For the pore of width $D=60\sigma$, we have continued the data collection for another two periods, each of $10^3\tau$ to test that the flow is stationary.

\paragraph*{Simulation software.}

The described NEMD simulations were conducted using \emph{HAL's MD package} \cite{HALMD,colberg2011}, a software that leverages the high parallelism of graphics processors (GPUs) and other accelerator hardware \cite{skoblin2023} and that is an efficient and precise tool for studying inhomogeneous fluids in complex setups \cite{hoefling2015,hoefling2024jcp,chaudhuri2016prb,ebrahimi-viand2020jcp,roy2024mp}.
A typical simulation run of the data collection phase of length $10^3 \tau$ (\num{e6} integration steps) for $D=15 \sigma$ (\num{e5} fluid particles), including the calculation of spatially resolved observables, takes approximately 8~minutes on a single Nvidia GPU of type GeForce RTX 2070.

\subsection{Observables}

\paragraph*{Density profile.}

For the spatial variation of the number density $\rho(z)$ perpendicular to the wall surfaces, we used a Fourier representation of the fluctuating density field, encoding all particle positions: $\hat \rho(z) = \sum_{j=1}^N \delta(z - z_j)$; here, it suffices to consider only the Cartesian component along the $z$-axis.
First, the (fluctuating) density modes \cite{Hansen:SimpleLiquids}:
\begin{equation}
 \hat\rho_{k_z}(t) = \sum_{j=1}^{N} \exp(i k_z z_j(t)) \,
 \label{eq:rho_kz}
\end{equation}
are computed for all wave ``vectors'' $k_z$ in parallel, by summing the contributions from all $N$ particles, positioned at $z_j(t)$ at time $t$;
these summations occur during the simulation and take full advantage of GPU acceleration.
The wave vectors are located on the reciprocal lattice of the periodic domain of the simulation;
here, $k_z = 2\pi n / L_z$ for $|n| \leq n_\text{max}$ integer and $L_z$ is the length of the simulation box in the $z$-direction.
The stationary density profile in real space, $\rho(z)$, follows from applying the corresponding Fourier backtransform to the time-averaged density modes:
\begin{equation}
  \rho(z) = L_z^{-1} \sum_{k_z} \langle \rho_{k_z}(t) \rangle \,\e^{-i k_z z - w^2 k^2} \,,
  \label{eq:rho-iFT}
\end{equation}
carried out once in a post-processing step.
The spatial resolution $\Delta z$ of the method is determined by the maximum wave number $k_\text{max} = 2\pi n_\text{max} / L_z$, which was chosen to be $n_\text{max}=120$ for $L_z=30\sigma$, corresponding to $\Delta z = \sigma/4$.
\Cref{eq:rho-iFT} includes a Gaussian filter of width $w=\Delta z/2$, which slightly smoothens the density profile $\rho(z)$ by softly truncating the spectrum at long wave numbers, thereby avoiding spurious small-wave length oscillations.

\paragraph*{Further spatially resolved properties.}

During data collection, phase space trajectories, combining particle positions and velocities, are saved regularly after intervals of length $\tau$ to the disk using the H5MD file format \cite{debuyl2014}.
Spatially resolved profiles of the flow velocity are calculated along the direction normal to the pore surfaces.
To this end, the pore space is divided along the $z$-axis into equally-spaced slabs of width $0.375\sigma$ and centered at the positions $z_n$.
For each of these ``bins'', the set $S_n(t)$ of particles located in bin $n$ at time $t$ is determined
and the mean velocity $v_x(z_n)$ is computed from:
\begin{equation}
v_x(z_n) =  \frac{1}{\langle |S_n(t)|\rangle} \Bigl\langle \sum_{j\in S_n(t)} u_{x,j}(t) \Bigr\rangle,
\end{equation}
where $|S_n(t)|$ represents the number of particles in the set $S_n(t)$, and $u_{x,j}(t)$ is the $x$-component of the velocity of particle $j$ at time $t$.

Similarly, spatially resolved profiles of the fluid's thermodynamic properties along the direction of the flow
are obtained by also dividing the simulation box into 40 slabs of width $2.5\sigma$ and perpendicular to the $x$-axis.
From the particles in each slab, instantaneous values of the particle count, potential energy, temperature, (virial) pressure, center-of-mass velocity, are calculated on-the-fly of the simulation using the GPU and are written to the above mentioned H5MD output file.
The ensemble averages $\langle \,\cdot\, \rangle $ are calculated in a post-processing step as the time average over the entire data collection phase.

\begin{figure}
    \centering
    \includegraphics[width=\figwidth,trim=50pt 20pt 0pt 70pt,clip]{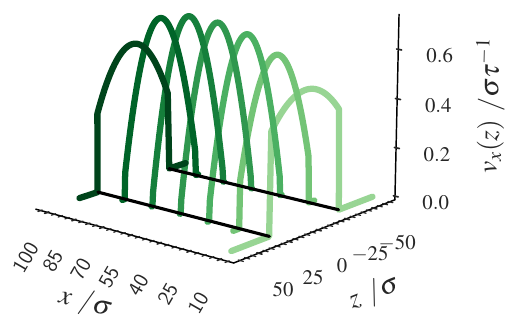}
    \caption{Exemplary transversal profiles of the fluid velocity $v_x(z)$ at different $x$ positions along the flow direction.
    The lines show parabolic fits to simulation data obtained for a wide slit-shaped pore ($D=60\sigma$) with amorphous walls.
    A mean velocity of $v_0=0.5\sigma\tau^{-1}$ is imposed on the fluid in the pump region, at the entrance of the pore ($x < 10\sigma$).
    One observes that a plug-like flow at the entrance develops into the classical parabolic shape of Poiseuille flow downstream.}
    \label{fig:velocity_profiles3d}
\end{figure}

The fluid particles leave the thermostatted pump region with a spatially homogeneous mean velocity, which does not depend on the distances to the pore surfaces.
This corresponds to the plug-like flow that is mathematically obtained for (almost) full-slip boundary conditions at the surfaces.
Only gradually, the plug flow transitions downstream into Poiseuille flow due to viscous stresses.
This transformation is exemplified in \cref{fig:velocity_profiles3d} for the flow inside a wide pore ($D=60\sigma$) and a pumping velocity of $v_0=0.5 \sigma \tau^{-1}$.
The transversal velocity profiles $v_x(z)$ at different $x$ positions exhibits an almost flat behavior near $x=L_\text{pump}$, with little velocity variation, which changes into a parabolic shape for $x \gtrsim 30 \sigma$
as the flow develops towards the center of the pore space.
This issue is less relevant for more narrow pores, \emph{i.e.}, the Poiseuille flow is already well developed closer to the pump region.
In the computation of quantities averaged along the flow, we take this effect into account by discarding the upstream part of the pore space and considering only particles for which $x > 30\sigma$; for the pore with $D=60\sigma$, only particles with $50 \sigma \leq x \leq 80 \sigma$ were considered.


\paragraph*{Shear viscosity.}

The dynamic shear viscosity $\eta_b$ of the bulk fluid was computed in equilibrium MD simulations \cite{roy2016} of the bulk fluid, which is homogeneous and isotropic, using the Einstein--Helfand formula \cite{helfand1, helfand2},
\begin{equation}
    \eta_b= \frac{\rho}{\kB T} \lim_{t \to\infty} \frac{1}{6N} \partial_t \sum_{\alpha<\beta}
    \Bigl\langle \Bigl(\int_0^t \! \Pi_{\alpha\beta}(t') \,\diff t'\Bigr)^2 \Bigr\rangle ,
    \label{eq:helfand_moment}
\end{equation}
for Helfand's second moment of the time-integrated, microscopic expression for the total stress tensor \cite{Hansen:SimpleLiquids},
\begin{equation}
    \tens\Pi(t) = \sum_{i=1}^N \Bigl[
        m \vec u_i(t) \otimes \vec u_i(t) +
        \sum_{j > i}^N \vec r_{ij}(t) \otimes \vec F_{ij}(t)
    \Bigr] \,,
\end{equation}
where $\vec u_i(t)$ is the velocity of particle $i$,
$\vec r_{ij}(t) = \vec r_i(t) - \vec r_j(t)$ the separation vector between particles $i$ and $j$
and, $\vec F_{ij}(t)$ the force acting on particle~$j$
due to particle~$i$, and $\otimes$ denotes the tensor product.
\Cref{tab:bulk-properties} lists the bulk values of the shear viscosity obtained for the repulsive fluid at different thermodynamics state points: for the quiescent fluid, for the largest pressure applied at the inlet, and for the highest temperature observed at the outlet.

With this, we can characterize the flow regime in terms of the Reynolds number:
\begin{equation}
    \text{Re} = \frac{m\rho \bar{v} D}{\eta_b},
\end{equation}
which is set by the mass density $m\rho$ of the fluid, the typical flow velocity $\bar{v} \approx v_0$, and the width $D$ of the slit pore;
the latter is supposed to be the typical length over which the fluid properties vary.
In this study, we shall consider pumping velocities $v_0$ between $0.1\sigma\tau^{-1}$ and $1\sigma\tau^{-1}$,
yielding to Reynolds numbers between $0.74$ and $7.4$.
This range of values suggests that the flow remains laminar and that it is sufficiently well described by the linearized Stokes equation.
Moreover, the investigated flows are still in the low Mach number regime: $\text{Ma} = v_0 / c_s \lesssim 0.2$ for the largest pump velocity used, see \cref{tab:bulk-properties}.
For comparison, employing the LJ parameters of argon, a velocity of $v_0=1 \sigma \tau^{-1}$ corresponds to \SI{160}{m\,s^{-1}}.
%
%

\section{Results \& Discussion}

\subsection{Surface layering in transversal density profiles}

\begin{figure}
    \centering
    \includegraphics[width=\figwidth]{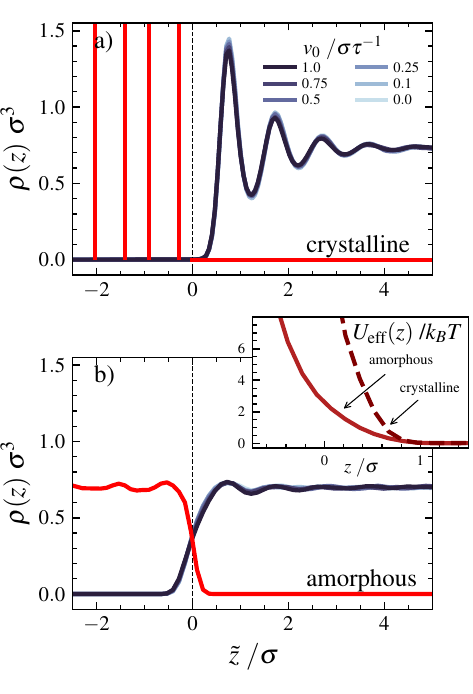}
    \caption{Number density $\rho(z)$ of the simulated fluid flow as a function of the distance $\tilde{z}=\left(z+D/2\right)$ to one of the pore surfaces (vertical, dashed lines); the surfaces stem from (a) crystalline and (b) amorphous walls.
    Black-blueish colors distinguish the results for six nominal pump velocities $v_0$, including the equilibrium case ($v_0=0$).
    Red lines depict the density profiles of the wall particles (``obstacles'') comprising the pore walls.
    For the crystalline walls (panel~a),
    the vertical lines indicate the positions of lattice planes.
    For amorphous walls (panel~b), obstacle configurations are such that their structure matches that of the fluid.
    The inset shows the effective potentials $U_\eff(z)$, summarizing the interaction of fluid particles with both wall structures that is obtained from averaging over the lateral position on the surfaces.}
    \label{fig:rho(z)}
\end{figure}

In a first step, we determined the implications of the morphology of the pore surfaces
on the fluid structure near the surfaces.
For the two kinds of surfaces, as realized by crystalline and amorphous walls, \cref{sec:wall-setup}, we have calculated
the number density profiles $\rho(\tilde z)$ as a function of the distance $\tilde z = z\pm D/2$ to the surfaces, \cref{fig:rho(z)}, both in equilibrium ($v_0=0$) and for fluid flows driven at nominal velocities between $v_0=0.1 \sigma\tau^{-1}$ and $1\sigma\tau^{-1}$.
For the calculation, only the $z$-component of the particle positions was used, \cref{eq:rho_kz}, and the profiles shown are thus averaged along the $x$- and $y$-axes, \emph{i.e.}, along the directions parallel to the surfaces.

A first observation is that the density profiles remain largely unaffected by changes in the flow velocity: even for the strongest driving studied, they closely reproduce the equilibrium profiles.
We have checked that the density profiles do not vary along the flow by computing the $\rho(\tilde z)$ locally, at different $x$ positions (data not shown).
This suggests that the flowing fluid is locally in equilibrium and that shear stresses due to the flow induce only minor, yet systematic changes in $\rho(\tilde z)$, which are visible near the density maxima.
It would be interesting to test this behavior near macroscopically curved surfaces, e.g., near a constriction of the pore space.

In case of the densely packed, crystalline walls, $\rho(\tilde z)$ displays three to four distinct monolayers of adsorbed fluid near the surfaces, which resemble the layering phenomena known for, \emph{e.g.}, hard spheres near a flat, impenetrable surface \cite{Henderson1984,evans2015jpcm}.
(As a side remark, such layering is absent in the composition profile for surface adsorption of binary fluids \cite{roy2024mp}.)
We interpret this behavior of $\rho(\tilde z)$ as a direct consequence of the geometric regularity and dense packing of the \emph{crystalline} obstacle configuration, shown in the simulation snapshot in \cref{fig:wall_setup}.

In contrast to this, the density profile lacks any pronounced peaks near the amorphous wall with the obstacle configuration matching the fluid structure, \cref{fig:rho(z)}b): $\rho(\tilde z)$ remains almost constant upon approaching the surface as close as $\tilde z \approx 1\sigma$ and, finally, decays to zero monotonically.
We conclude that the use of \emph{amorphous} walls in simulations can essentially eliminate the issue of layering effects, so that bulk-like behavior can be expected in a larger volume of the confined fluid.

Local equilibrium and the weak dependence of $\rho(\tilde z)$ on the flow velocity suggest further that the obtained density profiles can uniquely be associated with an effective surface potential \cite{evans1979,schmidt2022rmp}, $U_\eff(\tilde z)$.
We have obtained $U_\eff(\tilde z)$ within separate simulations from the equilibrium density $\propto \exp(-U_\eff(\tilde z) / \kB T)$ of an ideal gas filling the pore.
We find that $U_\eff(\tilde z)$ is a purely repulsive potential for both wall structures (inset of \cref{fig:rho(z)}),
which aligns with the construction of the pore walls from purely repulsive obstacles that are merely arranged differently.
However, $U_\eff(\tilde z)$ is considerably softer for the amorphous wall and this suffices to remove the layer structure from the fluid.

\subsection{Fluid depletion zone}
\label{sec:depletion}

The density profile near the crystalline wall, \cref{fig:rho(z)}a), indicates a depletion zone, reminiscent of a gaseous layer at the surface, which may be related to solvophobicity, but not necessarily, and which can lower the resistance to fluid flow, \emph{i.e.}, yield a large slip length.
The width $\ddepl$ of this depletion zone may be defined as follows via the density profiles $\rho(\tilde z)$ and $\rho_\text{obs}(\tilde z)$ of the fluid and the wall obstacles, respectively, which are compared to their bulk values $\rho_b$ and $\rho_{\text{obs},b}$ \cite{huang2008prl, janecek2007l}:
\begin{equation}
  \ddepl = \int_{-\infty}^{D/2} \left[
    1 - \frac{\rho_\text{obs}(\tilde z)}{\rho_{\text{obs},b}} - \frac{\rho(\tilde z)}{\rho_b}
  \right] \diff \tilde z \,;
  \label{eq:depletion-def}
\end{equation}
the coordinate $\tilde z = z + D/2$ is shifted such that the surface is at $\tilde z=0$.
Taking into account the finite width of the pore, the integral extends only across one of the pore surfaces, up to the pore center at $\tilde z=D/2$ and we use $\rho_b:=\rho(D/2)$ as the fluid bulk density.

For the amorphous wall configuration, the obstacles' density profile mirrors the fluid density, \cref{fig:rho(z)}(b);
hence, the depletion length is exactly zero, $\ddepl = 0$.
(Numerical integration of the simulated density profiles yields $\ddepl\approx 0.05\sigma$, close to zero.)
This reflects the complete absence of a depletion layer in the fluid--wall interface.

In the crystalline wall configuration, the obstacles are arranged periodically on an fcc lattice with a mean density of $\rho_{\text{obs},b} = 2.6 \sigma^{-3}$.
The corresponding density profile $\rho_\text{obs}(\tilde z)$ for $\tilde z < 0$ is a periodic sequence of Dirac $\ddepl$-peaks at the positions of the lattice planes, \cref{fig:rho(z)}(a).
Considering the integral over one unit cell, which contains two peaks, \emph{e.g.},
\begin{equation}
  \int_{-a_\text{lat}}^0 \left[ 1 - \frac{\rho_\text{obs}(\tilde z)}{\rho_{\text{obs}, b}} \right] \diff \tilde z = 0,
\end{equation}
shows that the obstacle contribution to $\ddepl$ vanishes.
In the pore space, $\rho_\text{obs}(\tilde z) = 0$ for $\tilde z > 0$ and for the simulated fluid density $\rho(\tilde z) = 0$ inside of the pore wall ($\tilde z < 0$). Hence, we can split the integral at the pore surface, leaving us with:
\begin{equation}
  \ddepl = \int_0^{D/2} \left[ 1 - \frac{\rho(\tilde z)}{\rho_b} \right] \diff \tilde z .
\end{equation}
The oscillations of the fluid density around $\rho_b$ contribute positively to $\ddepl$;
via numerical integration of $\rho(\tilde z)$, we obtain that the depletion layer between the fluid and the crystalline pore wall has a width of $\ddepl \approx 0.28\sigma$.

\begin{figure}
    \includegraphics[width=\figwidth]{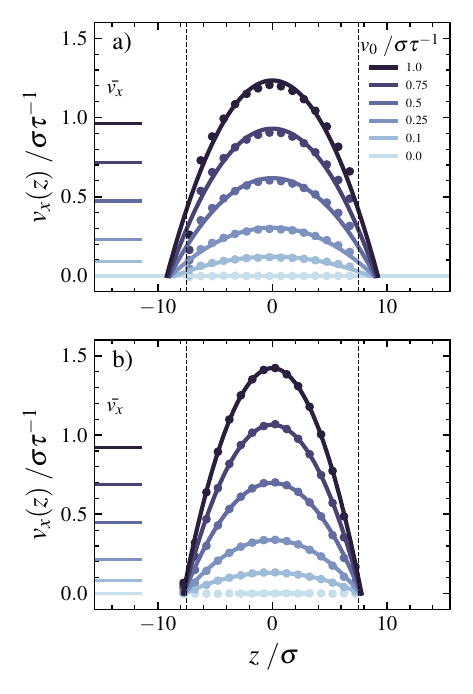}
    \caption{Variation of the flow velocity $v_x(z)$ transversal to the flow, as function of the distance $z$ to the pore center.
    The disks show NEMD simulation data for a slit pore of a nominal width of $D=15\sigma$ with (a) crystalline and (b) amorphous walls; for clarity, only every second data point is shown.
    Vertical, dashed lines indicate the positions $z=\pm D/2$ of the pore surfaces.
    Colors distinguish the results for six nominal pump velocities $v_0$; the realized average flow velocities $\bar v_x$ are indicated by the horizontal bars on the left.
    Solid lines are parabolic fits of the Poiseuille law to the data, allowing for partial slip boundary conditions, \cref{eq:HP-slip}.}
    \label{fig:HPfit}
\end{figure}

We emphasize that there is an ambiguity in the definition of the surface position. Here, we use $\tilde z=0$, \emph{i.e.}, we have \emph{ad hoc} set the distance between the surface and the first layer of the obstacle lattice to be $a_\text{lat}/4 \approx 0.3 \sigma$. This value stems from details of how the crystalline configuration was constructed, \cref{sec:wall-setup}.
However, a larger distance, \emph{e.g.}, $0.6 \sigma$, can be justified equally well, in particular, by noting that
$\rho(\tilde z = 0.3\sigma) = 0$ in the simulation data.
It implies that a shift $\delta z < a_\text{lat}/4$ of the nominal surface position towards the pore center directly subtracts from the previously calculated depletion length, $\ddepl - \delta z$, and also from the slip length.
The mathematical reason behind this ambiguity in $\ddepl$ is that the integral in \cref{eq:depletion-def} for a crystalline wall does not converge, but oscillates when sending the lower boundary to $-\infty$.
Hence, one could use $\ddepl = 0$ as criterion to \emph{define} the surface position by adjusting $\delta z$.
Under this convention, there is no depletion zone near the crystalline wall too.

\subsection{Poiseuille flow with slip}

\begin{figure}
    \includegraphics[width=\figwidth]{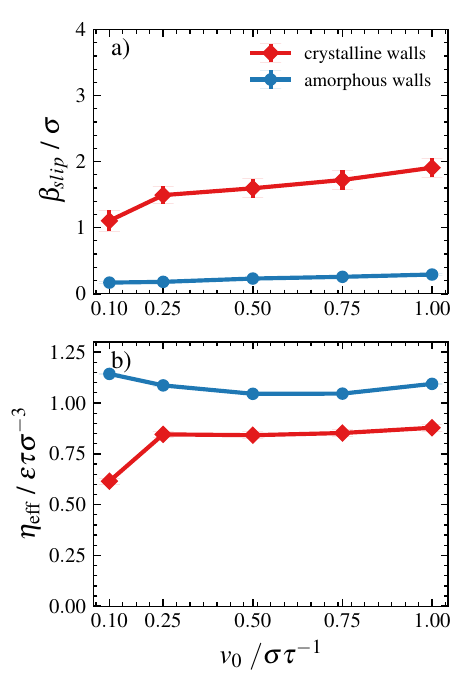}
    \caption{(a) Slip length $\bslip$ and (b) effective shear viscosity $\eta_\mathrm{eff}$ obtained from the fits to the velocity profiles in \cref{fig:HPfit} for Poiseuille flow between crystalline, red diamonds, and amorphous, blue disks, pore walls.
    Solid lines connect the data points and serve as guide for the eye.
    }
    \label{fig:HPfit_comparison}
\end{figure}

\begin{figure*}
    \centering
    \includegraphics[width=.33\linewidth]{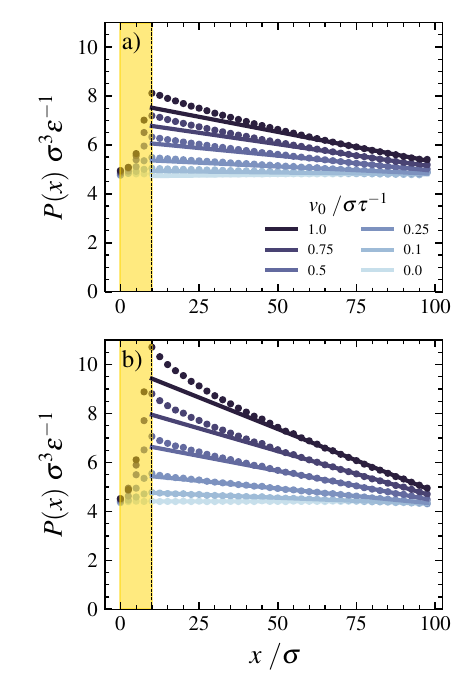}%
    \includegraphics[width=.33\linewidth]{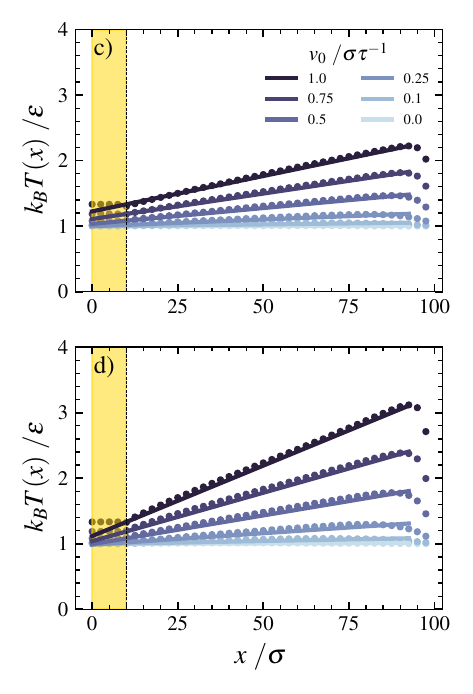}%
    \includegraphics[width=.33\linewidth]{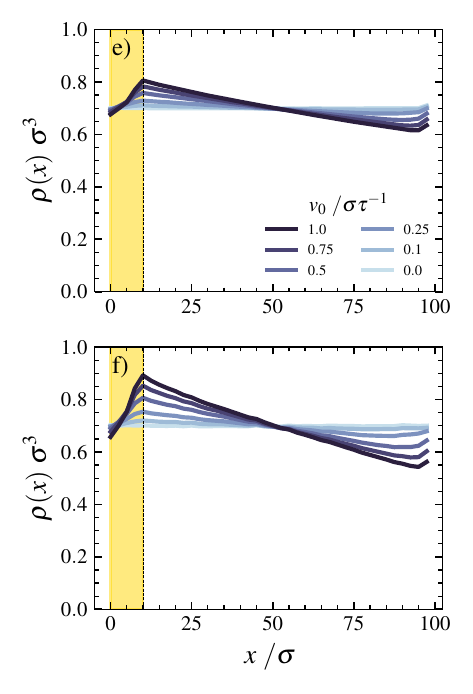}
    \caption{Variation of fluid pressure $P(x)$, temperature $T(x)$, and number density $\rho(x)$ along the direction of the flow inside a slit-shaped nanopore with either crystalline (top row) or amorphous (bottom row) walls.
    The pore is of width $D=15\sigma$ and has a length of $L=100\sigma$; periodic boundary conditions are applied along the $x$-axis.
    Disks represent simulation data and solid lines denote linear fits to the data for $x > 50\sigma$, indicating constant pressure and temperature gradients, respectively.
    Colors distinguish the six flow velocities $v_0$ imposed by a thermostatted pump in the pore entrance (yellow shaded regions).
    }
    \label{fig:profiles}
\end{figure*}

The surface morphology of the pore walls also determines the flow velocity $v_x(z)$ in the cross section of the pore, as a function of the distance to the pore center at $z=0$.
For the crystalline walls, the velocity profiles of the simulated fluid flows exhibit a parabolic shape in the pore space
for a range of pumping velocities $v_0$, \cref{fig:HPfit}a).
However, close to the surfaces, the obtained flow velocities are still considerably large,  approximately one half of the maximum value $v_x(0)$.
The data points next to the surface positions display jumps to values close to zero, which we attribute to the very low particle densities $\rho(z)$ at these positions and thus a poor statistics.
For all pumping velocities, the velocity profiles $v_x(z)$ are well described by the modified Poiseuille law, \cref{eq:HP-slip}, which permits also a non-zero slip length $\dslip$ or, equivalently $\bslip$.
The values of $\bslip$ fitted to the data depend on the strength of the fluid flow: they increase mildly from $\bslip \approx 1.10\sigma$ to $1.91\sigma$ upon increasing $v_0$ from $0.1\sigma \tau^{-1}$ to $1\sigma\tau^{-1}$, \cref{fig:HPfit_comparison}a).
This pronounced deviation from the no-slip boundary condition, expected to hold at macroscopic scales, suggests an only moderate resistance of the pore walls to the fluid's movement along the surfaces.
One possible contribution to the enhancement of the surface's flow is a putative sliding motion of the distinct layers that occur near the crystalline walls, \cref{fig:rho(z)}a).

In contrast to this, the amorphous walls yield nearly perfect no-slip boundary conditions for the flow, as shown in \cref{fig:HPfit}b).
The obtained slip lengths are an order of magnitude smaller than for the crystalline walls and increase mildly from $\bslip \approx 0.17\sigma$ to $0.29\sigma$ upon increasing the flow velocity, \cref{fig:HPfit_comparison}a).
Thus, fluid particles encounter a strong resistance as they flow past by the amorphous surface,
which we attribute to the disordered, fluid-like structure of this type of wall, generating an almost optimal roughness at the atomic scale with respect to minimizing the slip.
For such small slip lengths, we note that the value of $\beta_\text{slip}$ depends on the precise definition of the surface position.

\subsection{Effective viscosity of the confined fluids}

\begin{figure}
    \includegraphics[width=\figwidth]{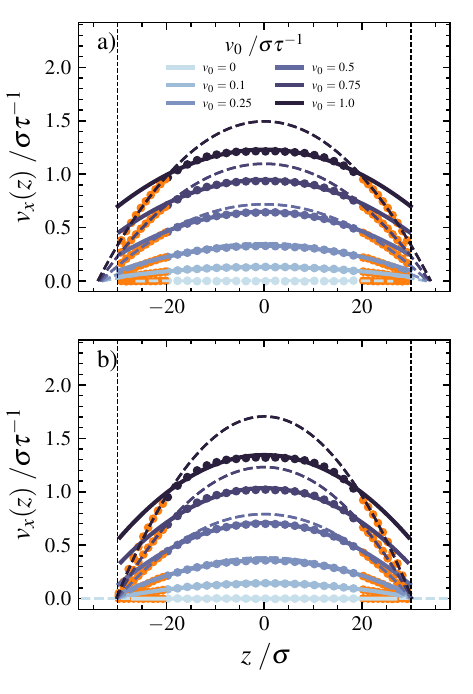}
    \caption{Flow velocity $v_x(z)$ transversal to the flow in the same representation as in \cref{fig:HPfit}, but for a pore of width $D=60\sigma$ with (a) crystalline and (b) amorphous walls.
    Solid lines are parabolic fits of the Poiseuille law for partial slip boundary conditions, \cref{eq:HP-slip}, to the central part of the flow, i.e., to the data spanning the region $-20 \leq z/\sigma \leq 20$.
    Dashed lines represent parabolic fits to the boundary regions of the flow, corresponding to data shown in orange.
    }
    \label{fig:HPfit_D60}
\end{figure}

For Poiseuille flow, the curvature of the velocity profile $v_x(z)$ equals the ratio of the pressure gradient $\Delta P/L$ and the shear viscosity $\eta$, and it does not depend on the boundary condition (\emph{i.e.}, the slip length $\bslip$), see \cref{eq:HP-slip}.
In order to test this relation in the NEMD simulations, we determined the fluid pressure $P(x)$ along the direction of the flow, \cref{fig:profiles}a)-b).
The pressure exhibits an approximately linear decrease due to the viscous resistance and the surface stresses imposed on the fluid as it moves through the pore.
For the pore with crystalline walls, \cref{fig:profiles}a), the pressure drop is less pronounced than in the case of amorphous walls, \cref{fig:profiles}b), which is inline with the picture of a lower resistance of the surfaces to the flow in the crystalline case.
The slope of the pressure profiles increases (in overall magnitude) as the imposed flow velocity is increased, suggesting that the stronger driving of the fluid leads to enhanced friction forces and also higher energy dissipation in the fluid.
A constant pressure gradient, $\partial_x P(x) \approx \Delta P/L$, is expected from the linearized Navier--Stokes equation for laminar and incompressible, fully developed flow in the absence of an external body force, see \cref{eq:PDE-HP}.
For fast flows and in the pore entrance ($x \lesssim 40 \sigma$), the profiles display deviations from a linear behavior of $P(x)$, which we attribute mainly to the transformation of plug flow to Poiseuille flow; compressibility effects of the simulated fluid appear to play a minor role, \cref{tab:bulk-properties}. Particles in this region were excluded from the data analysis.
Linear fits to the pressure data in \cref{fig:profiles} provided us with values for $\Delta P/L$ for each pumping velocity $v_0$ studied.

\begin{figure}
    \includegraphics[width=\figwidth]{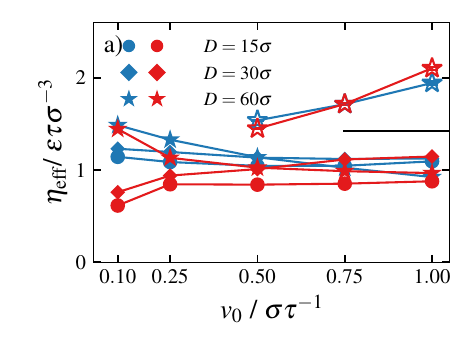}
    \hfill
    \includegraphics[width=\figwidth]{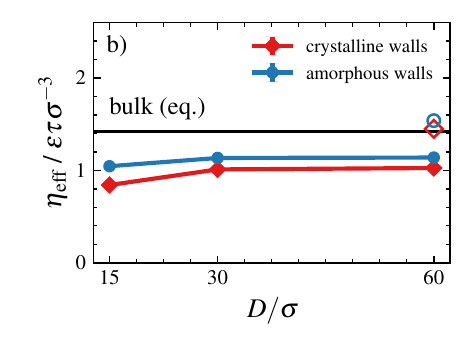}
    \caption{ a) Effective shear viscosity $\eta_\mathrm{eff}$ obtained from fits to the Poiseuille flow in \cref{fig:HPfit} as a function of the pumping velocity $v_0$ for three pore widths:
    $D = 15\sigma$ (disks), $D = 30 \sigma$ (diamonds), and $D = 60\sigma$ (stars).
    Colors distinguish results for pores with the crystalline, red, and amorphous, blue, wall configurations.
    Lines connect data points of the same pore to guide the eye.
    ~b)~Effective shear viscosity $\eta_\mathrm{eff}$ at a flow velocity of $v_0 = 0.5 \sigma \tau^{-1}$ as function of the pore width $D$ for the two surface morphologies.
    In both panels, open symbols represent data for $\eta_\eff$ obtained from the central region of the flow in the pore with $D=60\sigma$, \cref{fig:HPfit_D60}.
}
    \label{fig:viscosity_comparison}
\end{figure}

The simulated flows may be used to define an effective shear viscosity $\eta_\eff$ of the fluid as the ratio of $\Delta P/L$ and the curvature $\partial_x^2 v_x(z)/2$ of the parabolic velocity profiles in \cref{fig:HPfit}.
The viscosities obtained for $D=15\sigma$ depend only mildly on the flow velocity $v_0$, with an exception at the smallest value of $v_0$, \cref{fig:HPfit_comparison}b).
The values for the crystalline walls are slightly smaller than those for amorphous walls;
which is in line with observations in supercooled liquids that rough confinement can introduce additional friction and slows down structural relaxation \cite{scheidler2000epl,Scheidler2004}.
However, the overall magnitude $\eta_\eff \approx 1 \epsilon\tau \sigma^{-3}$ is well below the bulk value $\eta_b \approx 1.42\epsilon\tau \sigma^{-3}$, obtained in separate simulations of the homogeneous fluid under equilibrium conditions, \cref{tab:bulk-properties}.

In order to clarify the discrepancy between effective and bulk viscosities, we have performed NEMD simulations for fluid flow inside much wider slit-shaped pores, using $D=30\sigma$ and $60\sigma$, and have determined the effective viscosity for a range of pumping velocities.
We note that for the widest pore, $D=60\sigma$, and the fastest flows studied, $v_0 \gtrsim 0.5\sigma \tau^{-1}$,
the parabolic Poiseuille velocity profile is not yet fully developed: \cref{fig:HPfit_D60} shows a crossover between parabolas in the central part of the flow and in the boundary regions near the pore walls, despite the fact that sampling of the profiles was restricted to particles in the lower half of the conduit, at positions $50\sigma \leq x \leq 80\sigma$.
In these cases, we distinguish two values of $\eta_\eff$, determined from the curvature of $v_x(z)$ either in the central part (bluish disks) or in the boundary regions (orange disks);
the larger curvature of $v_x(z)$ in the boundary region yields a smaller effective viscosity.

The results for $\eta_\eff$ in the different pores (\cref{fig:viscosity_comparison}a) again show the dependency on the surface morphology and some variability for the different flow velocities $v_0$, similarly as in \cref{fig:HPfit_comparison}b).
Overall, $\eta_\eff$ tends to increase with the pore width $D$
and, if the values from the boundary layer are used for the widest pore, $\eta_\eff(D)$ converges for large $D$ but remains well below the bulk value $\eta_b$ of the quiescent fluid, \cref{fig:viscosity_comparison}b).
On the contrary, $\eta_\eff$ obtained in the central flow for $D=60\sigma$ increases markedly with the flow velocity and exceeds $\eta_b$ (open stars in \cref{fig:viscosity_comparison}a),
which we attribute to an increase of the viscosity due to elevated temperatures along strong flows, see \cref{fig:profiles,tab:bulk-properties}.
Our findings suggest that the viscosity of a structured, confined fluid differs from that of the corresponding homogeneous fluid,
in line with previous studies \cite{scheidler2000epl,Scheidler2004,hoang2012pre}, and the transition between the two cases is observed in the widest pore studied as a function of the distance to the wall.

\subsection{Permeability of the pore}

\begin{figure}
    \includegraphics[width=\figwidth]{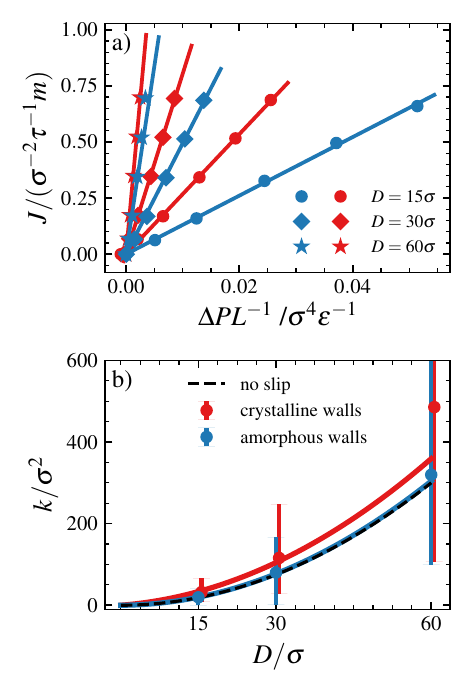}
    \caption{a) Relation between the mass flux $J_\rho$ imposed on the simulated fluid and the obtained pressure gradient $\Delta P/L$.
    Colors distinguish the two surface morphologies studied: crystalline walls (red) and amorphous walls (blue).
    Symbols encode the pore widths: $15 \sigma$ (circles), $30 \sigma$ (diamonds), and $60 \sigma$ (stars).
    Straight solid lines are fits to the data, testing Darcy's law, \cref{eq:Darcy-a}.
    ~b)~Permeability $k$ of the nanopore as a function of the pore width $D$. Symbols show the values obtained from the linear fits in panel~(a) for amorphous (blue disks) and crystalline (red disks) walls for each pore width.
    The symbols were slightly displaced horizontally for better visibility of the error bars, which reflect the uncertainty in the effective viscosity values (\cref{fig:viscosity_comparison}a), entering the calculation of $k$.
    Solid lines are weighted fits of \cref{eq:permeability_slip} to the data points.
    The black dashed line represents the permeability predicted for plane Poiseuille flow without slip, $k \propto D^2$.
    }
    \label{fig:permeability}
\end{figure}

The relation between the mass flux $J_\rho$ and the pressure gradient $\Delta P/L$ is shown in \cref{fig:permeability}a) for the three different pore widths analyzed, $D/\sigma = 15, 30, 60$.
The data exhibit a linear relationship for small pressure gradients, in agreement with Darcy’s law, \cref{eq:Darcy-a}.
Nonlinear behavior, \emph{i.e.}, a deviation from Darcy's law, begins to emerge at the largest pressure gradients obtained, particularly for the pore with $D=60\sigma$ and amorphous walls.
In general, the mass flux for the crystalline pore walls, red symbols, is consistently higher than for amorphous walls, blue symbols, across all pressure gradients, indicating a higher permeability $k$.
The latter was obtained from the slopes in \cref{fig:permeability}a) after dividing by the kinematic viscosity $\nu=\eta/(m\rho)$,
using the data from \cref{fig:viscosity_comparison}; this latter step introduces an appreciable uncertainty in the values of $k$, which is indicated in \cref{fig:permeability}b).
The observation of higher permeability for crystalline walls aligns with our findings on the slip length, \cref{fig:HPfit_comparison}a):
In Poiseuille flow through a slit-shaped conduit, the permeability is expected to increase quadratically with the width, $k =D^2/12$;
allowing for partial slip reduces the friction at the surfaces and leads to an increase of the permeability, depending on the slip length $\bslip$, \cref{eq:permeability_slip}.
Fitting this prediction to the uncertainty-weighted permeability data in \cref{fig:permeability}b) yields
$\bslip = (2.0 \pm 0.6) \sigma$ and $\bslip=(0.14 \pm 0.11) \sigma$,
for the crystalline and amorphous walls, respectively,
which is in good agreement with the more precise values obtained from the velocity profiles, \cref{fig:HPfit_comparison}a).
Overall, for amorphous pore walls, $\bslip / D \lesssim 0.02$ is negligibly small and the dependence of the permeability $k$ on the pore width $D$ is very well described by the no-slip law.

\subsection{Heating due to viscous dissipation}

The profiles of the fluid temperature $T(x)$ along the direction of the flow provide insights into the energy dissipation of the fluid as it is pushed through the nanopore.
We recall that the NEMD simulations are performed such that the total energy is locally conserved in the pore, so that the generated heat needs to be transported downstream to the thermostatted pump. In particular, the fluid flow is isolated energetically and a heat transfer into the pore walls is not included in the simulated set-up.

For both crystalline and amorphous walls, the obtained $T(x)$ increases approximately linearly with the flow, \cref{fig:profiles}c)-d).
For the smaller pumping velocities ($v_0 \lesssim 0.25\sigma\tau^{-1}$), the fluid is near global thermal equilibrium and the slight increase in temperature is consistent with a moderate energy dissipation.
However, at higher flow velocities ($v \gtrsim 0.5\sigma\tau^{-1}$), a pronounced rise of the temperature is observed, up to factors of 2 and 3, indicating a significant increase in heat generation, which we attribute to the higher frictional forces at play.
The temperature rise is weaker by about 30\% for the crystalline walls, \cref{fig:profiles}c), than for the amorphous ones, \cref{fig:profiles}d), which qualitatively aligns again with the increased surface slip in the first case.

The almost linear variation of pressure $P(x)$ and temperature $T(x)$ along the flow has consequences for the fluid density $\rho(x)$ since, anticipating local thermal equilibrium, the three quantities are coupled via the thermal equation of state.
The data exhibit also a linear change of the density for all simulated flows and wall configurations, \cref{fig:profiles}e)-f);
however, the overall variation of $\rho(x)$ is comparably small, on the order of 15--25\% for the largest flow velocity considered.
Hence, the fluid mechanical assumption of incompressible flow is approximately fulfilled; compressibility coefficients are given in \cref{tab:bulk-properties}

The observed linear increase of the temperature is supported by an argument from fluid mechanics, given above in \cref{sub:non-isothermal-flow}.
Starting from the basic conservation laws combined with the additional assumptions of an incompressible, Newtonian flow with uniform shear viscosity, one obtains Poiseuille flow profiles (with or without slip) on the one hand, but also a linear variation of the temperature along the flow, \cref{eq:grad-T-Poiseuille}.
Our simulation results in \cref{fig:profiles} corroborate the proportionality $\partial_x T \propto \partial_x P$ and are in semi-quantitative agreement with the prefactor predicted in \cref{eq:grad-T-Poiseuille};
the discrepancy is smaller for faster flows and varies between a few and up to 50 percent using the bulk properties of the quiescent fluid, \cref{tab:bulk-properties}.

Concerning a possible exponential growth of $T(x)$, which is also a potential solution to the non-isothermal flow equations, \cref{eq:grad-T-general}, we note first that the growth length $\ell = \kappa / (c_v J_\rho) \approx 6.6 \sigma^2 / (v_0 \tau)$
is approaching the length of the conduit for small and moderate flow velocities.
Second, the absence of such an exponential increase in our simulation results suggests that the initial slope $\partial_x T(0)$ is sufficiently close to the constant-gradient solution, $\bar\Phi/(c_v J_\rho)$.

%
%

\section{Summary and Conclusions}

In summary, we have investigated the role that the atomic-scale morphology of the pore surfaces plays in the flow behavior of a simple fluid confined to slit-shaped nanopores.
Utilizing boundary-driven NEMD simulations with large particle numbers, we examined two kinds of pore walls realized by crystalline and amorphous obstacle configurations, respectively;
in the latter case, the wall structure was chosen to match that of the equilibrium fluid.
We have shown that the two surface morphologies lead to very different surface resistances to fluid flow:
whereas a pronounced, well-known slip occurs in the flow near the crystalline walls,
high surface stresses near the amorphous walls drastically reduce the slip and lead to almost ideal no-slip boundary conditions (\cref{fig:HPfit}).
Accordingly, the obtained permeabilities of pores with crystalline walls are constantly higher than in the case of amorphous ones (\cref{fig:permeability}).

The simulated transversal velocity profiles $v_x(z)$, in the cross section of the pore, follow the predictions of Poiseuille theory for incompressible, Newtonian fluids, provided that a slip length is included, \cref{eq:HP-slip}.
The obtained slip lengths $\bslip$ are almost independent of the flow velocity and characterize the pore surface (\cref{fig:HPfit_comparison}).
However, the curvature of the parabolic velocity profiles suggests an effective shear viscosity $\eta_\eff$ that is significantly smaller than the bulk value $\eta_b$ of the quiescent fluid, $\eta_\eff \approx 0.7 \eta_b$.
We attribute this effect to the nanoconfinement, based on simulations for a range of pore widths. Our data provide evidence for a reduction of the viscosity in a boundary region near the pore surfaces (\cref{fig:viscosity_comparison})
and a crossover to the bulk value in the center of wide pores (\cref{fig:HPfit_D60}).

The fluid temperature increases linearly along the flow in the energy-conserving NEMD simulations, in particular, for fast flows (\cref{fig:profiles}c,d).
Such behavior is explained by the dominant role of viscous dissipation, quantified by $\bar\Phi$, in combination with fluid convection, given by the mass flux $J_\rho$.
Experimentally, heating by viscous dissipation was observed in strong Couette flows of dusty plasmas \cite{feng2012prl}.
Within the fluid mechanics of Newtonian fluids, \cref{eq:energyeqVII}, the viscous dissipation is treated as a localized heat source that increases the fluid temperature proportionally to the viscous stresses.
This leads to the prediction of a constant temperature gradient:
$\partial_x \bar T \propto \bar\Phi/J_\rho \propto \partial_x P$, see \cref{eq:grad-T,eq:grad-T-Poiseuille}.
The line of argument is consistent with the data for $T(x)$, showing smaller thermal gradients for the crystalline pore walls as compared to the amorphous walls.
The latter case implies a higher energy dissipation, which is attributed to the irregularity of the wall structure;
for crystalline walls, on the other hand, the comparably flat surface minimizes the frictional resistance to flow.
We acknowledge that an unbounded increase of the fluid temperature in the pore is against physical intuition and seems to contradict experimental evidence.
Yet, we note that (i) the temperature gradient is proportional to the pressure gradient, so the increase is not unbounded, (ii) measurements of the fluid temperature within a nanopore are not available to our knowledge, and (iii) heat transfer to and heat conduction within the pore material is not included in the investigated, oversimplified simulation model.
Such heat transfer obviously delimits the rise in temperature, depending on a set of additional parameters that one would need to introduce into the model along with additional non-trivial physical processes such as Kapitza heat resistance in the fluid--wall interface \cite{barrat2003mp,ge2006prl,auschra2021fp}.

Correlations between solvophobicity, fluid depletion, and slip length have been widely reported and questioned in the literature, based on experimental evidences, molecular simulations, and theoretical reasoning \cite{neto2005rpp, Israelachvili2011, Lau2003, cottin-bizonne2003nm, bocquet_2010, huang2008prl, barrat1999prl, bocquet_2007, evans2015jpcm, evans2017jcp, willard2014jcp, tocci2014nl, thiemann2022acs}.
More generally, one may ask whether and how fluid structure and kinetics are linked.
The notion of solvophobicity rests on the definition of the three-phase contact angle of a sessile droplet on the surface.
The fluid model investigated here uses the purely repulsive WCA pair potential and does not form droplets in equilibrium.
Hence, the usual definitions of the contact angle via Young's equation or the droplet shape \cite{weijs2011pf} are not applicable
and, for the present study, the concept of solvophobicity cannot be filled with a meaning.

At the nanoscale, a different angle on a fluid's surface behavior is provided by the transversal density profile $\rho(z)$, which can exhibit a depletion zone \cite{huang2008prl, janecek2007l}.
The simulated fluid shows strong density variations and layering near the crystalline walls (\cref{fig:rho(z)}a);
whereas the fluid remains homogeneous at comparable distances to the amorphous walls, which is accompanied by a monotone decay of the density to zero in the fluid--wall interface (\cref{fig:rho(z)}b).
For both wall types, we observe that the fluid structure is almost independent of the flow velocity and resembles the equilibrium behavior, also at the fastest flows studied.
A depletion length $\ddepl$ can be defined in terms of the density profiles of fluid and wall particles, \cref{sec:depletion}.
There is no depletion for the structure-matched, amorphous wall, $\ddepl=0$, where we observe almost no-slip behavior.
For the crystalline wall, with developed slip, we find that $\ddepl$ ranges between $\approx 0$ and $\approx 0.58\sigma \approx a_\text{lat} /2$ due to an ambiguity in the definition of the surface position for such walls; the choice $\ddepl = 0$ appears not unreasonable and may be used to locate the surface.
While the concept of a depletion zone proved fruitful for other systems to explain fluid slippage near surfaces \cite{huang2008prl}, its universality is challenged by the fluid--wall interface considered here.


The robustness of the density profiles with respect to flow suggests that the fluid can locally be treated as in equilibrium (in a co-moving frame)
and that there is a unique effective surface potential $U_\eff(z)$ seen by the fluid, which generates the density profile $\rho(z)$.
We have shown that changing the surface morphology can significantly alter the fluid structure and thus $U_\eff(z)$.
For the amorphous wall considered here, $U_\eff(z)$ is a repulsive potential and sufficiently soft to remove the layer structure from the fluid (\ref{fig:rho(z)}).
Moreover, replacing the pore walls by flat, planar surfaces with interaction potential $U_\eff(z)$ would create the same density profile as near a structured wall while eliminating the surface friction to the flow, yielding plug flow and full-slip boundary conditions.

It was pointed out previously \cite{evans2015jpcm} that there is no simple correlation between the wetting properties of a fluid and the surface potential (and thus the fluid structure).
Similarly, we have provided a simple example for the absence of surface slip that is not intuitively inferred from an effective surface potential.
Rather than focusing on the dependence on a single variable $\tilde z$, measuring the distance to the surface, we propose that the three-dimensional structure of the fluid--wall interface, encoded in inhomogeneous two-point correlation functions \cite{evans1979}, must be considered to clarify possible relations between surface slip and fluid structure.
Such correlations can, in principle, be probed in grazing-incidence x-ray diffraction \cite{dietrich1995pr,hoefling2024jcp}.
For theoretical research, the further analysis of fluid behavior near amorphous, fluid-matched wall structures is a potentially fruitful endeavor.
Our findings point to the importance of molecular-level details in designing nanoporous materials for specific fluid transport applications. Tailoring the surface interactions and their morphology in nanoporous materials has the potential to optimize the efficiency of nanoscale filtration, drug delivery, or energy storage.

\appendix

\section{Derivation of an equation for the temperature field}
\label{sec:temperature-equation}

We start from the first law of thermodynamics applied to a small control volume $V$.
The total energy $E$ within the volume consists of the internal energy due to molecular interactions and thermal motion, and the kinetic energy due to the macroscopic fluid flow, but it does not include the potential energy due to external forces.
According to the first law, the rate of change $\diff E/\diff t$ of the energy balances the rate $\dot Q$ of heat added to the volume minus the rate $\dot W$ of work performed by the volume:
\begin{equation}
    \diff E/\diff t = \dot{Q} - \dot{W} \,.
    \label{eq:Ilawthermo-a}
\end{equation}
The energy content of the control volume changes due to local energy conversion and due to advection with the flow field:
\begin{multline}
    \frac{dE}{dt} =\int_V  \partial_t \bigl(m\rho e + \tfrac{1}{2} m\rho |\vec v|^2\bigr) \, \diff V \\
    + \int_{\partial V} \bigl(m\rho e + \tfrac{1}{2} m\rho |\vec v|^2\bigr) \vec v \cdot \unitvec n \,\diff A ,
\end{multline}
where the field $e$ denotes the internal energy per unit mass, $m\rho$ is the mass density field, and $\vec v$ the flow field, as above.
The work performed by the volume has a contribution stemming from the pressure acting on its surface $\partial V$ and a contribution due to viscous forces at the surface. For the rate of work:
\begin{equation}
    \dot{W} = \int_{\partial V} P \vec v \cdot \unitvec n \, \diff A
        - \int_{\partial V} \vec v \cdot (\tens\tau \cdot \unitvec n) \, \diff A \,,
\end{equation}
where the unit vector $\unitvec n$ is the outward normal to the surface and the symmetric tensor $\tens\tau$ is the deviatoric stress of the fluid.
Moving from the rate of heat, $\dot Q$, to that of the heat density, per volume, $\dot q$, and
transforming the surface integrals into volume integrals via Gauss' divergence theorem yields
a local form of energy conservation:
\begin{multline}
    \partial_t \bigl(m\rho e + \tfrac{1}{2} m\rho |\vec v|^2\bigr)
    + \nabla \cdot \bigr[m\rho \vec v \bigl(e+ P/(m\rho) + \tfrac{1}{2} |\vec v|^2\bigl)\bigl] \\
    = \dot q + \nabla \cdot (\vec v \cdot \tens\tau).
    \label{eq:energy-cons-a}
\end{multline}
The \emph{l.h.s.}\ is further simplified by expanding the derivatives over the products and
introducing the material derivative, $\Dt=\partial_t +\vec v \cdot \nabla$,
as well as by making use of the continuity equation for mass conservation,
\begin{equation}
    \partial_t  (m\rho) + \nabla \cdot (m\rho \vec v) = \Dt m\rho + m\rho \nabla \cdot \vec v = 0 \,
    \label{eq:mass-cons-a}
\end{equation}
and Cauchy's momentum equation:
\begin{equation}
    m\rho \Dt \vec v = -\nabla P + \nabla \cdot \tens\tau \,.
    \label{eq:momentum-cons-a}
\end{equation}
On the \emph{r.h.s.} of \cref{eq:energy-cons-a}, we rewrite
$\nabla \cdot (\vec v \cdot \tens\tau) = \vec v \cdot (\nabla \cdot \tens\tau) +  \Phi$
upon defining the viscous dissipation function:
\begin{equation}
    \Phi = \tens\tau : \nabla \vec v = \tau_{\alpha\beta} (\partial_\alpha v_\beta) \,;
    \label{eq:Phi-def-a}
\end{equation}
the sum convention is applied in the last expression.
Collecting terms leads to the energy equation, which holds generally, here, in the absence of external forces, since it is merely an expression of the fundamental conservation laws:
\begin{equation}
    m\rho \Dt e + P \nabla \cdot \vec v
    = \dot q + \Phi \,.
    \label{eq:Ilawthermo_IV-a}
\end{equation}

Assuming thermodynamic equilibrium locally, in each control volume, spatial changes of the internal energy $e$ are determined by the fluid's caloric and thermal equations of state, $e=e(T,\rho)$ and $P=P(T,\rho)$.
We invoke the fundamental relation \cite{clausius1879, Callen1985}:
\begin{equation}
  \diff e(T,\rho) = c_v(T,\rho) \, \diff T - P(T,\rho) \, \diff (\rho^{-1}) \,,
\end{equation}
where the irreversibly added heat $\delta q = c_v \diff T$ is expressed through the isochoric specific heat $c_v = (\partial e/\partial T)_\rho$ and $\rho^{-1}$ denotes the specific volume.
Then, the chain rule for $\Dt$ implies:
\begin{align}
 \Dt e(T,m\rho)
    &= \left(\frac{\partial e}{\partial T}\right)_{\!\rho} \Dt T
    + \left(\frac{\partial e}{\partial \rho}\right)_{\!T} \Dt m\rho \notag \\
    &= c_v \Dt T + m\rho^{-2} P \,\Dt \rho \,.
\end{align}
Thus, the \emph{l.h.s.}\ of \cref{eq:Ilawthermo_IV-a} can be rewritten to yield the temperature form of the energy equation:
\begin{equation}
    c_v m\rho \Dt T = \dot q + \Phi \,,
    \label{eq:temperature-eq-a}
\end{equation}
where we have used mass conservation, \cref{eq:mass-cons-a}, to cancel the term $P \nabla \cdot \vec v$.
In particular, the thermal equation of state is not needed to describe the evolution of the temperature field.

For the rate $\dot Q$ of change of the heat contents of the control volume, we consider only heat conduction through the control surface, which is given in terms of the heat flux vector $\vec j_q$ by $\dot Q = -\int_{\partial V} \vec j_q \cdot \unitvec n \, dA$.
Using Gauss' theorem and removing the integral over the control volume, which was arbitrary, $\dot q = - \nabla \cdot \vec j_q$ holds.
We apply Fourier's law \cite{fourier1822, Jiji2024},
\begin{equation}
    \vec j_q=-\kappa \nabla T,
\end{equation}
which approximates the heat flux $\vec j_q$ as the linear response to a (small) temperature gradient $\nabla T$,
in accordance with experimental observations for both fluids and solids;
the coefficient $\kappa$ is the thermal conductivity.
Hence,
\begin{equation}
   \dot q  = \kappa \nabla^2 T.
   \label{eq:heat-equation-a}
\end{equation}

For the viscous dissipation function, \cref{eq:Phi-def-a}, we specialize to an incompressible, Newtonian fluid. Under these conditions the viscous stress tensor is $\tens\tau = \eta \dot{\tens\gamma}$, where the rate of strain tensor is given by $\dot{\tens\gamma} = \nabla \vec v + (\nabla \vec v)^\top$.
Accordingly, the viscous dissipation function reads:
\begin{align}
  \Phi &= \eta (\partial_\alpha v_\beta + \partial_\beta v_\alpha) (\partial_\alpha v_\beta)
  = \tfrac{1}{2} \eta \dot{\tens\gamma} : \dot{\tens\gamma} \,,
  \label{eq:Phi-incompressible-a}
\end{align}
and substituting into \cref{eq:temperature-eq-a} yields to the equation of the temperature field under the named approximations \cite{winter1987}:
\begin{equation}
    c_v m\rho \Dt T = \kappa \nabla^2 T + \tfrac{1}{2} \eta \dot{\tens\gamma} : \dot{\tens\gamma} \,.
    \label{eq:temperature-eq-viscous-a}
\end{equation}

\section*{Data availability statement}

The data that support the findings of this study are available from the corresponding author upon reasonable request.

\begin{acknowledgments}
Financial support by Deutsche Forschungsgemeinschaft (DFG, German Research Foundation) under Project No.\ 523950429 is gratefully acknowledged.
\end{acknowledgments}

\bibliography{references}

\end{document}